\documentclass[aip,jcp,preprint,onecolumn,groupedaddress]{revtex4-1}
\usepackage{amsmath}	
\usepackage{setspace}	
\usepackage{graphicx}	
\usepackage{subfigure}	
\usepackage{color}		
\usepackage{verbatim}	
\graphicspath{{./figs/}}

\begin{document}

\title{ Suppressing the Rebound of Impacting Droplets from Solvophobic Surfaces by Polymer Additives: Polymer Adsorption and Molecular Mechanisms}


\author{Eunsang Lee}
\affiliation{Eduard-Zintl-Institut f\"ur Anorganische und Physikalische Chemie, Technische Universit\"at Darmstadt, Alarich-Weiss-Stra{\ss}e 8, 64287 Darmstadt, Germany}
\email{e.lee@theo.chemie.tu-darmstadt.de}
\author{Hari Krishna Chilukoti}
\affiliation{Eduard-Zintl-Institut f\"ur Anorganische und Physikalische Chemie, Technische Universit\"at Darmstadt, Alarich-Weiss-Stra{\ss}e 8, 64287 Darmstadt, Germany}
\altaffiliation{Current affiliation: Department of Mechanical Engineering, National Institute of Technology Warangal, Warangal, Telangana 506004, India}
\author{Florian M\"uller-Plathe}
\affiliation{Eduard-Zintl-Institut f\"ur Anorganische und Physikalische Chemie, Technische Universit\"at Darmstadt, Alarich-Weiss-Stra{\ss}e 8, 64287 Darmstadt, Germany}


\date{\today}
\begin{abstract}

A liquid droplet impacting on a solvophobic surface normally rebounds.
The rebound is suppressed by a small amount of dissolved polymer. 
In this work, using multi-body dissipative particle dynamics simulations, two anti-rebound mechanisms, the slow-retraction and the slow-hopping mechanisms, are identified.
Which of them dominates depends on the polymer-surface attraction strength.
However, these two mechanisms are not excluding each other but may coexist. 
During the droplet rebound, the surface-adsorbed polymer acts in two ways: the adsorbed beads mediate solvent-surface interactions, and highly stretching unadsorbed polymer segment exerts a retraction force on the liquid.
Both actions increase the friction against retraction and the resistance against hopping. 
We also investigate the effects of the molecular weight and the concentration of the polymer additive, the droplet size, and the impact velocity on the rebound tendency.
As the first work to provide a microscopic explanation of the anti-rebound mechanism by polymer additives, this study allows better understanding of wetting behavior by polymer-solution droplets.

\end{abstract}


\maketitle


\section{Introduction}

Understanding the physics of wetting of a solid surface by a liquid droplet is of importance in industry, agriculture, bio-engineering, or even in our everyday life. 
In last few decades, attention has been drawn to static and dynamic wetting by Newtonian droplets.\cite{dussan1979, de1985, werder2003, bonn2009, zhang2014}
A droplet impacting on a solid flat surface can splash, rebound, or deposit depending on the surface wettability, liquid inertia, viscosity, surface tension of the droplet, and many other factors.\cite{josserand2016} 
The balance between above factors, typically represented in terms of dimensionless numbers such as Weber (We) and Reynolds numbers (Re), predict or explain reasonably well the impact fate of Newtonian droplets.\cite{pasandidehfard1996,josserand2016,ukiwe2005,attane2007}
In particular, the balance between surface wettability and liquid inertia determines the tendency to rebound, \cite{mao1997, bartolo2006, yokoi2009} as a water droplet is very likely to rebound on a superhydrophobic surface.
Since the rebounding droplets are problematic in many practical applications, \textit{e.g.}, spraying, ink-jet printing, and coating and painting in industrial processes, controlling the droplet rebound is both a necessity and a great challenge.

Since a few hundred ppm of polyethylene oxide (PEO) were found to suppress the droplet rebound from a hydrophobic surface in 1990s, polymers have been widely accepted as anti-rebound agents.\cite{wirth1991, monteux2008, bertola2013}
This phenomenon attracted even more scientific attention, because it could not be explained in terms of classical dimensionless numbers.
After a droplet hits a solvophobic surface, the rebound involves three phases. 
The first is spreading during which the droplet loses its spherical shape and laterally expands into a pancake. 
The liquid's surface tension drives the second phase, retraction, in which the droplet nearly recovers sphericity. 
Finally, the lateral influx is converted into perpendicular momentum, and in this hopping stage, the droplet moves upward until it detaches from the surface.

The early successful technique to provide the underlying physical picture was high-speed imaging.\cite{bergeron2000}
This work argued that a droplet of dilute PEO aqueous solution has a lower retraction velocity than a pure water droplet, and the slow retraction is caused by the non-Newtonian elongational viscosity of the PEO solution.
They showed that the retraction velocity can be written in terms of the capillary number for both pure-water and for polymer-solution droplets if the capillary number is calculated not using the shear viscosity but the elongational viscosity.
This argument, however, has been debated due to the unclear definition of the elongational flow during the retraction stage.\cite{bertola2013}
The reduced retraction velocity was blamed as a source of the anti-rebound also by other research groups since the observation of a polymer deposit on the substrate left behind the receding contact line.\cite{smith2010, bertola2013, huh2015}
In this work, the stick-slip dynamics of the spreading diameter relates the polymer contribution to the additional friction on the liquid-solid interface. 
The reduced retraction velocity explained by the sliding angle of droplets with PEO and silica nano-particles proposed basically the same idea of the increased contact line friction,\cite{zang2013, zang2014} but a molecular picture is still missing.
More recently, the time scale of polymer stretching in comparison with the shear rate during retraction was examined in an experiment of a droplet of aqueous PEO impacting on a Teflon surface with varying velocity.\cite{dhar2019}
This work suggested that, in order to suppress the rebound by the elongation force, the shear rate achieved by the initial impact velocity should be faster than the polymer relaxation.
Yet another viewpoint for the anti-rebound was suggested by an experiment which showed a high damping of the height oscillations after the retraction by adding PEO to the aqueous droplet.\cite{chen2018}
This result implies that the elongation force is possibly acting not only during retraction, but also after the retraction.

Despite the experiments, many questions still remain open because the impact process is extremely non-equilibrium and not even close to a steady state. 
This makes it very difficult to find a relation between the rebound tendency and any transport properties of the liquid or solution.
Simulations would be useful to get an intuition of the anti-rebound mechanism, but only a few field-based simulation studies for impacting droplets with polymer additives with a Finite-Element-Nonliner-Elastic-Chilcott-Rallison model (FENE-CR) have been published so far. \cite{izbassarov2016,wang2017,tembely2019} 
In these works, however, the rebound outcome of polymer solution droplets was attributed to the change of the solution viscosity, which is, however, insignificant for the dilute polymer solutions.
Furthermore, one of the studies showed that a polymer-solution droplet with a higher viscosity than a pure solvent droplet shows a bigger tendency to rebound, which implies that the viscosity is not the origin of the rebound suppression.\cite{izbassarov2016}

More promising for getting a microscopic picture of the polymer-induced anti-rebound mechanism are particle-based simulations, since they can uncover the underlying molecular process.
Recently, we published the, to our best knowledge, first such simulation of an impacting droplet of a dilute polymer solution aided by multi-body dissipative particle dynamics (MDPD).\cite{lee2020}
We confirmed the rebound suppression after the retraction stage, namely, at the hopping stage, with a certain choice of material (simulation parameters).
The polymer adsorbed on the surface plays two roles in the rebound suppression. 
Firstly, it changes the effective wettability of the surface by mediating unfavorable solvent-surface interaction.
Secondly, it becomes highly stretched  perpendicular to the surface at the hopping stage, and it reels in the droplet as it recovers its coiled equilibrium configuration. 
Both contributions were quantified by the number of polymer beads adsorbed on the surface, which directly indicates adsorptivity. 
For a given surface adsorptivity (same number of adsorbed polymer beads), a droplet containing longer polymer chains is less likely to rebound.
Thus, longer polymer impedes better the rebound by offering larger resistance to the droplet hopping.
In that work,  we focused on the anti-rebound in a specific limited parameter range.
Simulations for a wider range of parameters are still necessary to establish whether these anti-rebound mechanisms are the only ones.

We discuss molecular mechanisms of rebound suppression for a wide parameter range.
We still confine our interest to the practically most relevant case at a dilute polymer in a good solvent whose Weber and Reynolds numbers, We and Re, are not much different from those of a pure solvent droplet.
In the first part, we will show the anti-rebound mechanism for different polymer-surface attraction strength.
After discussing the effect of the polymer concentration and the molecular weight on the rebound, we will briefly explain the droplet-size effect.
The last part of Results and Discussion contains the rebound tendency depending on the impact velocity.

\section{Simulation Method and Model}

\subsection{Multi-body Dissipative Particle Dynamics}

We used MDPD simulations developed to allow vapor-liquid coexistence to be simulated by an additional attractive nonbonded force to investigate dynamic droplet wetting.\cite{warren2003, espanol1995, espanol2017, jamali2015}
MDPD simulation employs three types of interaction forces, conservative ($F_{ij}^\mathrm{C}$), dissipative ($F_{ij}^\mathrm{D}$), and random forces ($F_{ij}^\mathrm{R}$).
The conservative force is the sum of pairwise force given by: 
\begin{equation}\label{eqn-dpdcons}
F_{ij}^\text{C}=B_{ij}w^\mathrm{B}(r_{ij})\cdot\bm{e}_{ij} + A_{ij}(\bar{\rho}_i +\bar{\rho}_j)w^\mathrm{A}(r_{ij})\cdot\bm{e}_{ij}\text{,}
\end{equation}
where $A_\text{ij}$ and $B_\text{ij}$ are parameters for the interaction between $i$ and $j$ beads, and $r_{ij}$ is  the distance between the two beads.
$w^{\text{B}}(r_{ij})$ is a cutoff function which linearly decreases with $r_{ij}$ and vanishes for $r>r_\text{B}$ : thus $w^{\text{B}}(r_{ij})=1-r/r_\text{B}$. 
$w^\text{A}(r_{ij})$ has the same form, but a different cutoff distance $r_\text{A}$. 
In this work, we use $r_\text{A}=$ 0.75 and $r_\text{B}=$ 1.0. 
The key feature of MDPD is the second attractive term, which depends on the local particle density $\bar{\rho}_i$:
\begin{equation}\label{eqn-localdens}
\bar{\rho}_i=\frac{15}{2\pi r_\text{B}^3} \sum_{j\neq i}w^{\text{B}}(r_{ij})\text.
\end{equation}
The dissipative and the random forces are taken from the regular dissipative-particle-dynamics (DPD) simulation written by:
\begin{align}
F_{ij}^\text{D}=-\gamma_\mathrm{D} w_\text{R}^2(r_{ij})\bm{v}_{ij}\cdot\bm{e}_{ij}\text,\\
F_{ij}^\text{R}=\sigma_\mathrm{R} w_\text{R}(r_{ij})\theta_{ij} \cdot\bm{e}_{ij}\text,
\end{align}
respectively, where $w_\text{R}(r_{ij})$ indicates the cutoff function with the cutoff distance $r_\text{R}=r_\text{A}=$ 0.75.
$\theta_{ij}$ is a Gaussian random variable.
Here, $\sigma_\text{R}$ and $\gamma_\mathrm{D}$ are the amplitude of the thermal fluctuation and the strength of the dissipative force, respectively, which are correlated with each other by the fluctuation-dissipation theorem  $\sigma_\mathrm{R}^2/2\gamma_\mathrm{D}=k_\text{B}T$.
In this work, all variables and parameters are expressed in terms of DPD reduced units.
The unit length, the unit mass, and the unit energy are $\sigma$, $m$, and $\epsilon$, respectively.
This defines the time unit as $\tau=(m\sigma^2/\epsilon)^{1/2}$.
Therefore, $A$ has an unit of $\epsilon\sigma^2$ and $B$ has an unit of $\epsilon\sigma^{-1}$. 
$\gamma_\mathrm{D}$ and $\sigma_\mathrm{R}$ have units of $\epsilon\tau\sigma^{-2}$ and $\epsilon\tau^{1/2}\sigma^{-1}$, respectively.
Throughout the paper, we omit the units for simplicity.
We use in this simulation $\gamma_\mathrm{D}=4.5$, $k_\mathrm{B}T=1.0$ and as a result, $\sigma_\mathrm{R}=3$.
Modified Velocity-Verlet algorithm with an integration time step $\Delta t$= 0.02 is employed for position and velocity integrations.\cite{groot1997}
In this method, because the DPD force is dependent on the velocity we need to predict a velocity at time $t+\Delta t$ by using the force at time $t$, which is later updated by a force at time $t+\Delta t$. 
The mass of all particles in our simulations is 1.

\subsection{Model Description}

The system of the impacting droplet simulation includes three types of beads: solvent (S), polymer (P), and surface (``wall'') beads (W).
The parameters for the repulsive force are same for all pair types, such that the amplitude and the cutoff distance are given as $B=25$ and $r_B$=0.75.
The cutoff distances of the attractive force for all pair types are also identical to 1.0.
We choose the amplitude of the attractive force to mimic a good solvent at the simulation temperature $k_\mathrm{B}T=1$, such that $A_\mathrm{P/P}=A_\mathrm{S/S}=A_\mathrm{S/P}=-40$.
The attraction amplitude between solvent and the surface is fixed to -10, which gives the static contact angle ($\cos \theta_0$) of the pure solvent droplet of 155\textdegree.(See our former study\cite{lee2020} for the calculation method)
Depending on the attraction amplitude between the polymer and the surface beads ($A_\mathrm{P/W}=-10$, ... , -200), different anti-rebound mechanisms are achieved.
Linear polymer is modeled with beads connected by harmonic springs with an equilibrium bond length of 0.65 and a force constant of 300.
The static properties of the polymer solution given in the SI confirms the good-solvent conditions for the chosen parameters.
The surface is composed of frozen surface beads on a square lattice, and the bounce-back reflection boundary condition is used.\cite{zhao2017} 
Gravitation is ignored as it is negligible.

The initial configuration before impact is set up from a static droplet configuration. 
The static droplet is generated by first pulling all particles toward the center of the box by a harmonic potential. 
It is then equilibrated  without the restraint for a time longer than five times of the longest polymer relaxation time.
The equilibrium density and the diameter of the drop are obtained by fitting the radial particle density with a hyperbolic tangent function. 
The liquid-vapor surface tension of the droplet is obtained by the Irving-Kirkwood method, but the pressure tensor is transformed onto the radial coordinate with respect to the center-of-mass of the droplet.\cite{ghoufi2010}
Then the difference between radial and tangential pressure tensor components is integrated.\cite{ollila2009}
The surface tensions obtained for different polymer concentrations and lengths seem to be almost the same within the statistical error.(See SI)
A separate bulk simulation of each polymer solution also gives the zero-shear viscosity, which also turns out to be the same within the statistical error.
As a result, the dimensionless numbers at the given impact velocity of 2 do not vary very much with the polymer composition, such that Re=50--62 and and We=105--110.
The details about the droplet's equilibrium properties are given in SI.

\section{Results and Discussion}

\subsection{Molecular Mechanisms of Rebound Suppression}

\begin{figure*}[t!]
\centering
\includegraphics[width=1.0\textwidth]{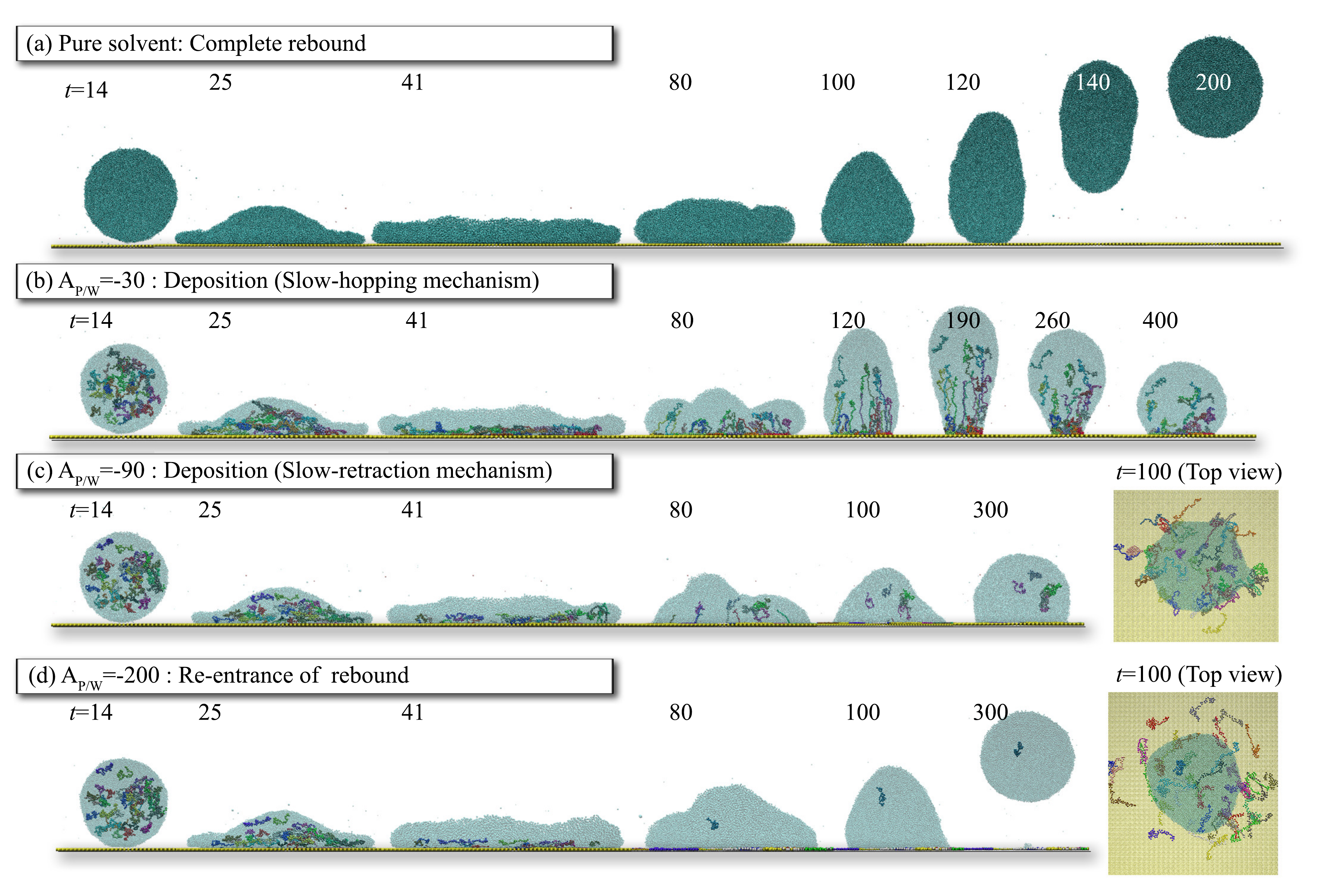}
\caption{\label{fig-outcomes} Simulation snapshots of (a) pure solvent, (b) polymer solution droplets at $A_\mathrm{P/W}=-30$, (c) at $A_\mathrm{P/W}=-90$, and (d) at $A_\mathrm{P/W}=-200$. A solvent particle is colored in cyan and each polymer chain is depicted in different color. The rightmost figures in Figures (c) and (d) show the top view of the droplets at $t=100$ (during the hopping stage). }
\end{figure*}

\begin{figure*}[h!]
\centering
\includegraphics[width=1.0\textwidth]{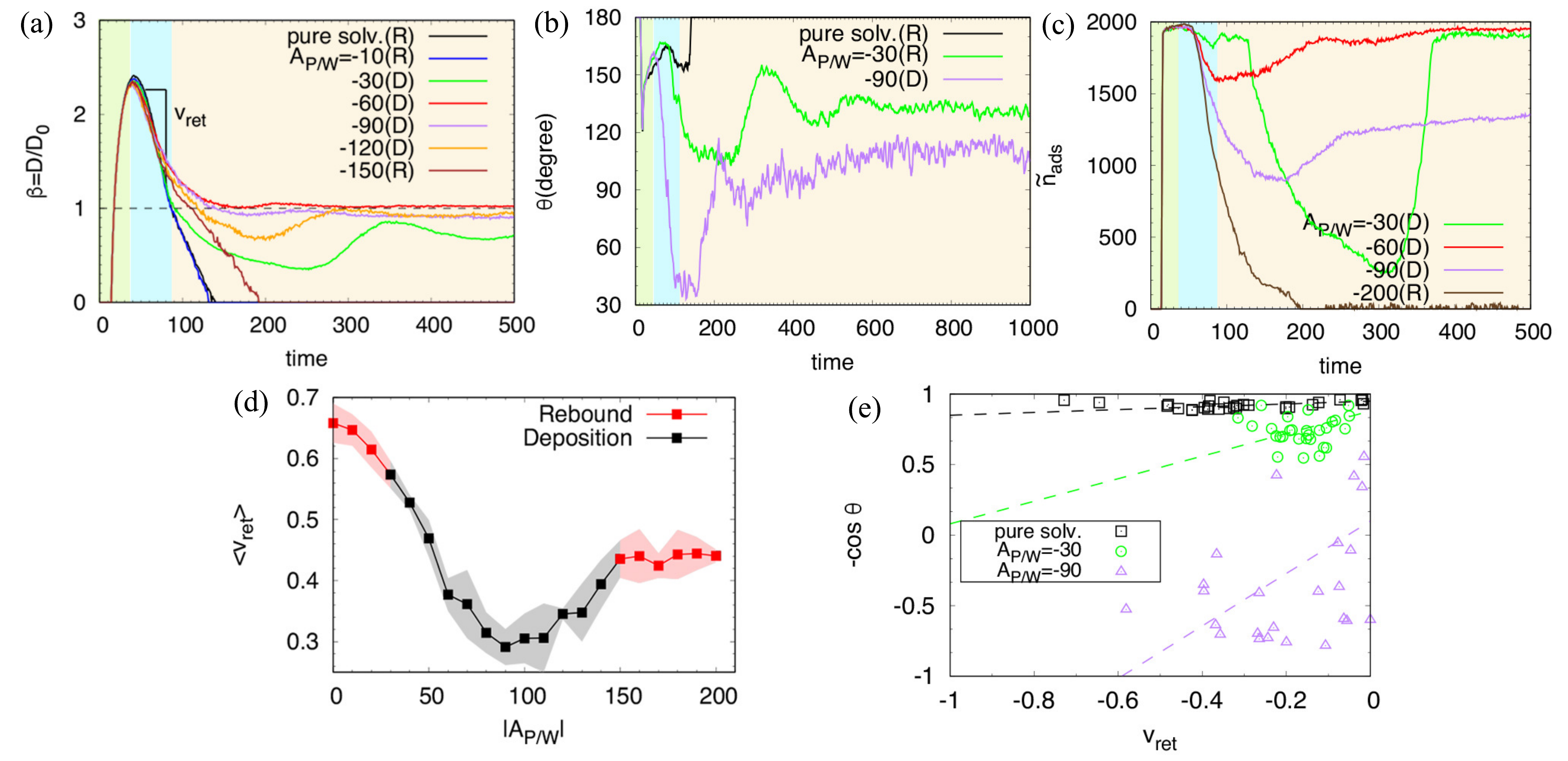}
\caption{\label{fig-mechanism} (a) The spreading factor, (b) the contact angle, and (c) the number of adsorbed polymer beads on the liquid-solid interface (\textit{i.e.}, inside the contact area) as a function of time for different values of $A_\mathrm{P/W}$. In figures (a),(b) and (c), the letter in the parenthesis of the legend indicates a (R) rebounding or (D) deposited droplet. The shaded regions by light green, light blue, and light yellow indicate the spreading, the retraction, and the hopping stages, respectively. (d) The average retraction velocity as a function of $|A_\mathrm{P/W}|$. The shaded area shows the error calculated from five trajectories for each point. (e) The contact angle as a function of the retraction velocity which are sampled in every 50 integration time steps during the retraction. Here, the retraction velocity is calculated by differentiating $D$ with respect to time with the finite difference method. Data points for each $A_\mathrm{P/W}$ are fitted by a line represented by the dashed line. }
\end{figure*}

Several droplet impact experiments have been interpreted by the ``\textit{slow-retraction}'' mechanism, in which the polymer deposited on the surface increases the friction between liquid and solid at phase 2, the retraction stage, i.e., polymer is suspected of slowing the horizontal motion of fluid, assuming that the surface is horizontally oriented.\cite{bergeron2000, monteux2008, smith2010, bertola2013, bertola2015, huh2015, chen2018, dhar2019}
On the other hand, our former study\cite{lee2020} and the experiment by Chen \textit{et al.}\cite{chen2018}  found that the reduced rebound velocity can be explained by the resistance against the hopping motion.
In this so-called ``\textit{slow-hopping}'' mechanism, the polymer obstructs the vertical motion of the fluid.
Both anti-rebound mechanisms are attributed to the strong polymer-surface attraction. 
The slow-retraction mechanism obviously appears as a consequence of the polymer-surface attraction being stronger than the polymer-solvent attraction.
It is also manifest in the slow-hopping mechanism by anchoring the droplet to the surface.
In the computer simulation, we study the rebound tendency and its mechanism as the polymer-surface attraction strength ($A_{\mathrm{P/W}}$) is varied from -10 to -200.
For this simulation, a fixed polymer length $N_\mathrm{p}$=50 (number of DPD beads in a polymer) and concentration $x_\mathrm{p}$=0.02 (the number fraction of polymer beads) are used. 
The diameter of the static free droplet of the polymer solution is 31.5 and the surface tension is 7.14.
With the impact velocity of 2, the dimensionless numbers are obtained as Re=53 and We=107, which are not much different from those of the pure solvent droplet, Re=57, We=106.

Figure \ref{fig-outcomes} shows the simulation snapshots of impacting droplets of the pure solvent and the polymer solution at different $A_\mathrm{P/W}$. 
The anti-rebound mechanism is also clearly indicated by the spreading factor, $\beta=D/D_0$, in Figure \ref{fig-mechanism}(a), where $D$ and $D_0$ refer to the diameter of a circle formed by the three-phase contact line and the initial droplet diameter, respectively.
The pure solvent droplet completely rebounds, as designed by the chosen simulation parameter (Figure \ref{fig-outcomes}(a)).
For $A_\mathrm{P/W}=-30$, the time taken for the retraction is almost the same as for the pure solvent droplet. 
This means that the retraction velocity is not much slower than for the pure solvent droplet, and the main resistance against the rebound appears only during the hopping stage at $t>90$ when the droplet tries to hop from the surface.
One can find in Figure \ref{fig-outcomes}(b) that many polymer beads are adsorbed on the surface and that the unadsorbed parts of the polymer are highly stretched in the hopping direction at $t>90$.
Therefore, the rebound is suppressed mainly by the slow-hopping mechanism.

When $A_\mathrm{P/W}$ further decreases to -90, the mechanism changes to slow-retraction. 
In Figure \ref{fig-mechanism}(a), while the spreading process and the resulting maximum spreading factor do not significantly differ from other values of $A_\mathrm{P/W}$, the contact line retracts only slowly, and $\beta$ does not pass below unity at the end of retraction. 
A hopping motion is hardly discriminable because particles have little momentum in the hopping direction after the retraction. 
The retraction velocity calculated from the linear fit of $\beta$ \textit{versus} $t$ during the retraction stage in Figure \ref{fig-mechanism}(d) shows a substantial reduction at $A_\mathrm{P/W}=-90$ compared to $A_\mathrm{P/W}=0$.
The contact angle ($\theta$) for $A_\mathrm{P/W}=-90$ in Figure \ref{fig-mechanism}(b) also supports the slow-retraction mechanism, as $\theta$ decreases to 30\textdegree~ and slowly approaches to the static value from below.
The corresponding droplet shape with the small contact angle is found at time $t=80$ and 100 in Figure \ref{fig-outcomes}(c).
The simulation snapshot from the top at this time in the same figure shows that the polymer molecules are largely adsorbed on the surface especially on the three-phase contact line, which induces an additional friction between the retracting contact line and the surface.
Simultaneously, its other chain end is not adsorbed but lies on the liquid-gas interface as shown at $t=80$ in Figure \ref{fig-outcomes}(c).
This dangling end is highly stretched and moves along with the solvent flow during retraction.
Consequently, the stretched polymer exerts an elongation force on the receding contact line, which contributes to reducing the retraction velocity and the receding contact angle.
The fact that $\beta$, $\theta$ and even the droplet shapes at $A_\mathrm{A/P}=-90$ are very well in agreement with the experimental results of the aqueous PEO solution droplet\cite{smith2010, bertola2013, bertola2015} indicates that the slow-retraction mechanism is caused by the strong polymer-surface attraction.

The molecular kinetic theory of wetting\cite{blake1969, blake2006} explains the effective friction on the moving contact line by the relation between the velocity of a moving contact line, $v_\text{CL}$, and the dynamic contact angle, $\theta$:
\begin{equation}
v_\text{CL}=2\kappa_0\lambda\sinh\Big(\frac{\gamma(\cos\theta_0-\cos\theta)}{2nk_\mathrm{B}T}\Big)\text,
\label{eqn-mkt}
\end{equation}
where $\kappa_0$, $\lambda$, $n$ and $\gamma$ refer to the frequency of the molecular hopping to the nearest adsorption site, the average length between nearest adsorption sites, the number of adsorption sites per unit area, and the liquid-vapor interfacial tension, respectively.
Equation \ref{eqn-mkt} can be reduced for states close to equilibrium as $v_\text{CL}\approx\gamma(\cos \theta_0-\cos \theta)/\zeta_{CL}$ with an assumption of $n\approx\lambda^{-2}$.
Here, $\zeta_{CL}=k_\mathrm{B}T/\kappa_0 \lambda^3$ is the coefficient of the effective friction on the three phase contact line.
Therefore, for a large friction on the moving contact line, $-\cos \theta$ increases rapidly with increasing the retraction velocity, and \textit{vice versa}.
Since the retraction velocity and the dynamic contact angle are not steady during the retraction, we scatter-plot the dynamic contact angle as a function of the retraction velocity sampled in every 50 integration steps during the retraction in Figure \ref{fig-mechanism}(e). 
In this figure, the slope of the fitted line increases with $A_\mathrm{P/W}$, indicating an increased effective friction on the contact line.

Decreasing $A_\mathrm{P/W}$ even further shows an important physical origin of the anti-rebound mechanism.
Figures \ref{fig-mechanism}(a) and \ref{fig-mechanism}(d) show that the very strong polymer-solvent attraction strength of $A_\mathrm{P/W}\leq -150$ results again in a droplet rebound. 
The reason for the re-entrance of rebound is that a very strongly surface-adsorbed polymer loses its capacity to act against hopping, while its resistance to retraction flow has reached saturation and does not increase further. 
The retraction velocity at $A_\mathrm{P/W}\leq -150$ is as slow as at $A_\mathrm{P/W}\approx -50$.
Yet, the slow-retraction is not enough to suppress the rebound.
This shows that the resistance against the hopping motion still plays an important role to suppress the rebound even when the retraction is already slow. 
In other words, the slow-hopping and the slow-retraction mechanisms are not exclusive, but complementary. 

Evidence for this is found in the snapshot of the droplets from the top at $t=100$ for $A_\mathrm{P/W}=-90$ and $-200$ (rightmost of Figures \ref{fig-outcomes}(c) and \ref{fig-outcomes}(d)). 
At $A_\mathrm{P/W}=-200$, the polymer deposits irreversibly on the surface.
It does not move together with the retracting solvent, but is left behind the receding contact line.
While the polymer molecules moving together with the receding contact line at $|\mathrm{P/W}|<90$ accumulate on the contact line, the irreversibly deposited polymer cannot contribute anymore either to the friction of the contact line or to the resistance against the hopping. 
As the number of polymer beads adsorbed on the surface ($n_\mathrm{ads}$) is the most relevant molecular determinant of the rebound tendency,\cite{lee2020} we calculate the number of surface-adsorbed beads, specifically, located inside the circular three-phase contact line ($\tilde{n}_\mathrm{ads}$) in Figure \ref{fig-mechanism}(c).
We thus disregard surface-adsorbed polymer outside the liquid-solid contact area.
Here, a bead whose distance from the surface is less than unity, the range of the attraction, is regarded as an adsorbed bead.
During retraction, it decreases more rapidly at $A_\mathrm{P/W}=-200$ than at $A_\mathrm{P/W}=-90$ (Figure \ref{fig-mechanism}(c)).
Therefore, despite the stronger polymer-surface attraction strength, the droplet at $A_\mathrm{P/W}=-200$ has a smaller friction of the contact line, and hence, a larger retraction velocity than at $A_\mathrm{P/W}=-90$.
Moreover, the very strong polymer-surface attraction leaves few polymer chains on the liquid-solid interface at the beginning of hopping.
Together, both effects lead to weak resistance against hopping and the re-entrance of rebound.

To summarize, the molecular origin of the anti-rebound by polymer additives is (i) the increased wettability due to the adsorbed polymer and (ii) the polymer elongation force, both of which lead to the slow-retraction and slow-hopping mechanisms. 
The weighting between the mechanisms is determined by the polymer-surface attraction strength, but the two mechanisms are not completely distinguishable and occur together.

\subsection{Concentration and Molecular Weight of Polymer}

\begin{figure*}[t!]
\centering
\includegraphics[width=0.8\textwidth]{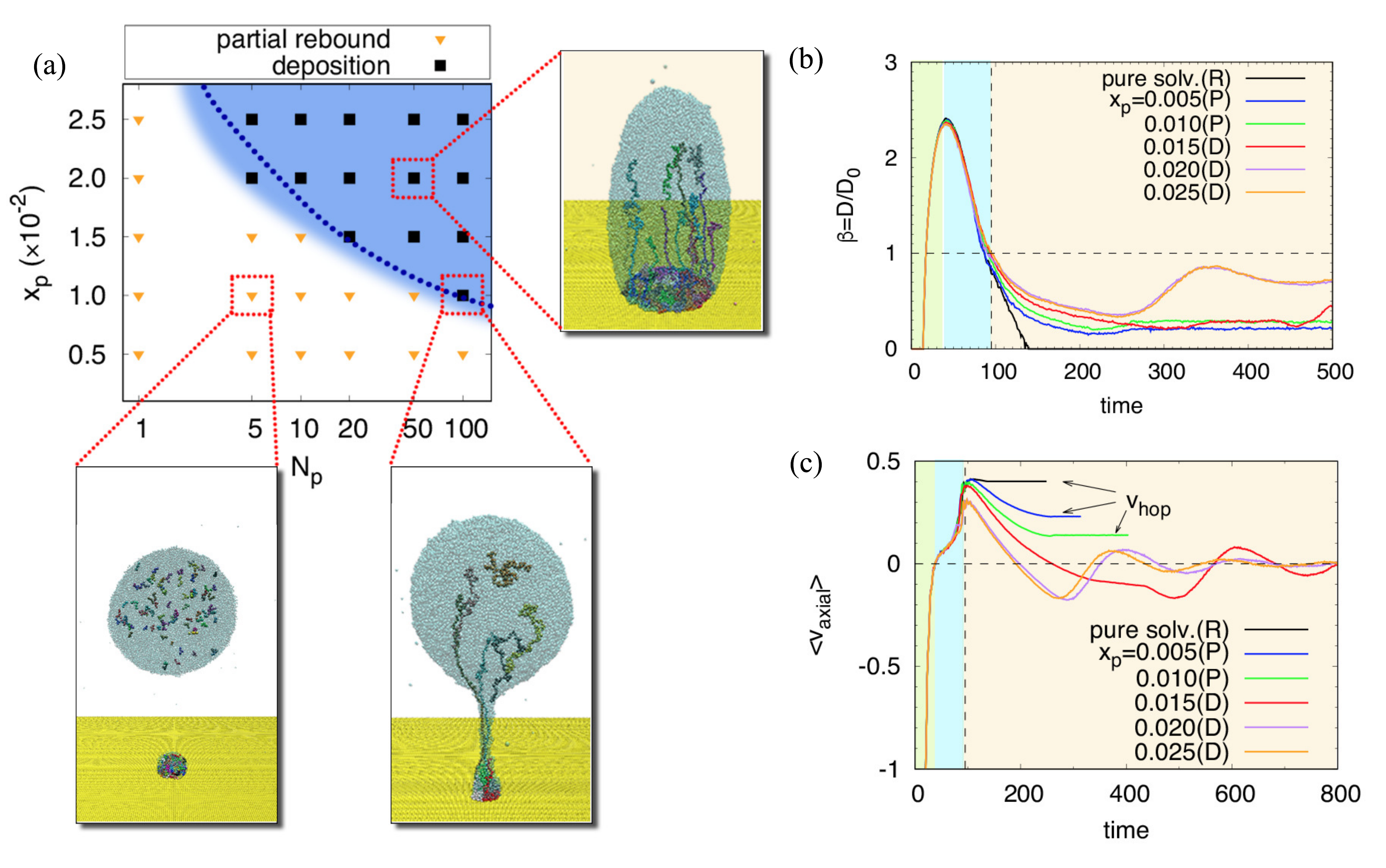}
\caption{\label{fig-polcomp} (a) Most probable outcome of rebound at different polymer concentrations and lengths. The blue shaded area indicates an anti-rebound regime. The blue dashed line indicates the boundary for the small droplet with $N=4\times 10^4$ in our former work.\cite{lee2020} The lower left and the lower right simulation snapshots show the rebounding droplet with the partial drop left on the surface and the polymer deposition with a long polymer neck (filament) during the hopping stage, respectively. The snapshot on the right shows highly stretched polymer conformation during the hopping stage of the deposited droplet. (b) The spreading factor and (c) the average axial velocity of droplet particles as a function of time for $N_\mathrm{p}$=50 at different concentrations. The shaded regions by light green, light blue, and light yellow indicate the spreading, retraction, and hopping stage, respectively. In (c) the constant velocity after hopping defines the hopping velocity $v_\mathrm{hop}$.}
\end{figure*}

A more detailed picture of the anti-rebound mechanism can be obtained by considering the polymer concentration and molecular weight dependence on the anti-rebound.
We performed impact simulations of a droplet with $10^5$ particles, which is larger than our former study.\cite{lee2020}
To restrict the simulation parameter space, we used the polymer-surface attraction strength of $A_\mathrm{P/W}=-30$ and the impact velocity of 2.
This choice gives Re$\approx 56$ and We$\approx 106$, which are in a reasonable experimental range. 
It also allows us to investigate the effect of droplet size by comparing with the former study (See the next section).
Static droplets have almost the same Re and We irrespective of the polymer composition.(See SI)
We find three characteristic outcomes of droplet impact: complete rebound, partial rebound, and deposition.(The process involving polymer necking conformation is not separated from deposition.)
Figure \ref{fig-polcomp}(a) shows the most probable outcome of rebound at different polymer lengths ($N_\mathrm{p}$) and concentrations ($x_\mathrm{p}$). 
Here, the partial rebound refers to a rebounding droplet with a small satellite droplet left on the surface (the lower left  snapshot of Figure \ref{fig-polcomp}(a)). 
Consistent with the former study, both a longer polymer and a higher polymer concentration reduce the rebound. 
We also find that droplets near the rebound-deposition boundary transiently show a very long neck between the surface and the temporary droplet above the surface (the lower right snapshot of Figure \ref{fig-polcomp}(a)). 
The long thin liquid filament in this morphology has also been observed in many experiments on non-Newtonian liquids,\citep{wagner2005, ingremeau2013, keshavarz2016, sachdev2016, chen2018, song2019} and strongly suggests the importance of the elongation force during the hopping stage.
Especially, the work of Song \textit{et al.} in which the polymer additive limits the fragmentation of a droplet when it impacts on a wired surface, is perfectly in line with the role of the polymer elongation force counteracting rebound.\cite{song2019}

Figure \ref{fig-polcomp}(b) shows the spreading factor as a function of time for droplets of different polymer concentrations for $N_\mathrm{p}=50$. 
In this figure, the maximum spreading factor ($\beta_\mathrm{max}$) does not depend on the polymer composition, which is consistent with other studies.\cite{bergeron2000, bertola2013, huh2015}
The main difference appears in the hopping stage, which supports an increasing importance of the slow-hopping mechanism.
Figure \ref{fig-polcomp}(c) shows the average axial velocity of droplet particles as a function of time.
In this figure, the reduction rate of $\langle v_\mathrm{axial}\rangle$, corresponding to the force opposing hopping, also decreases with increasing $x_\mathrm{p}$.
Therefore, the resistance against hopping is stronger for either larger $x_\mathrm{p}$ or longer $N_\mathrm{p}$, which leads to the rebound suppression.
However, in Figure \ref{fig-polcomp}(b), the retraction velocities (slope of $\beta$) also seem to be slightly different from each other, which was not observed in our former study with smaller droplets of $D_0=23.2$.
The perpendicular velocity, $\langle v_\mathrm{axial}\rangle$, at the beginning of the hopping stage (vertical dashed line at $t=90$) also decreases with increasing $x_\mathrm{p}$ in Figure \ref{fig-polcomp}(c), thus the anti-rebound is achieved not only by the slow-hopping but also by the slow-retraction mechanism.
The influence of droplet size is discussed below.

\begin{figure*}[t!]
\centering
\includegraphics[width=1.0\textwidth]{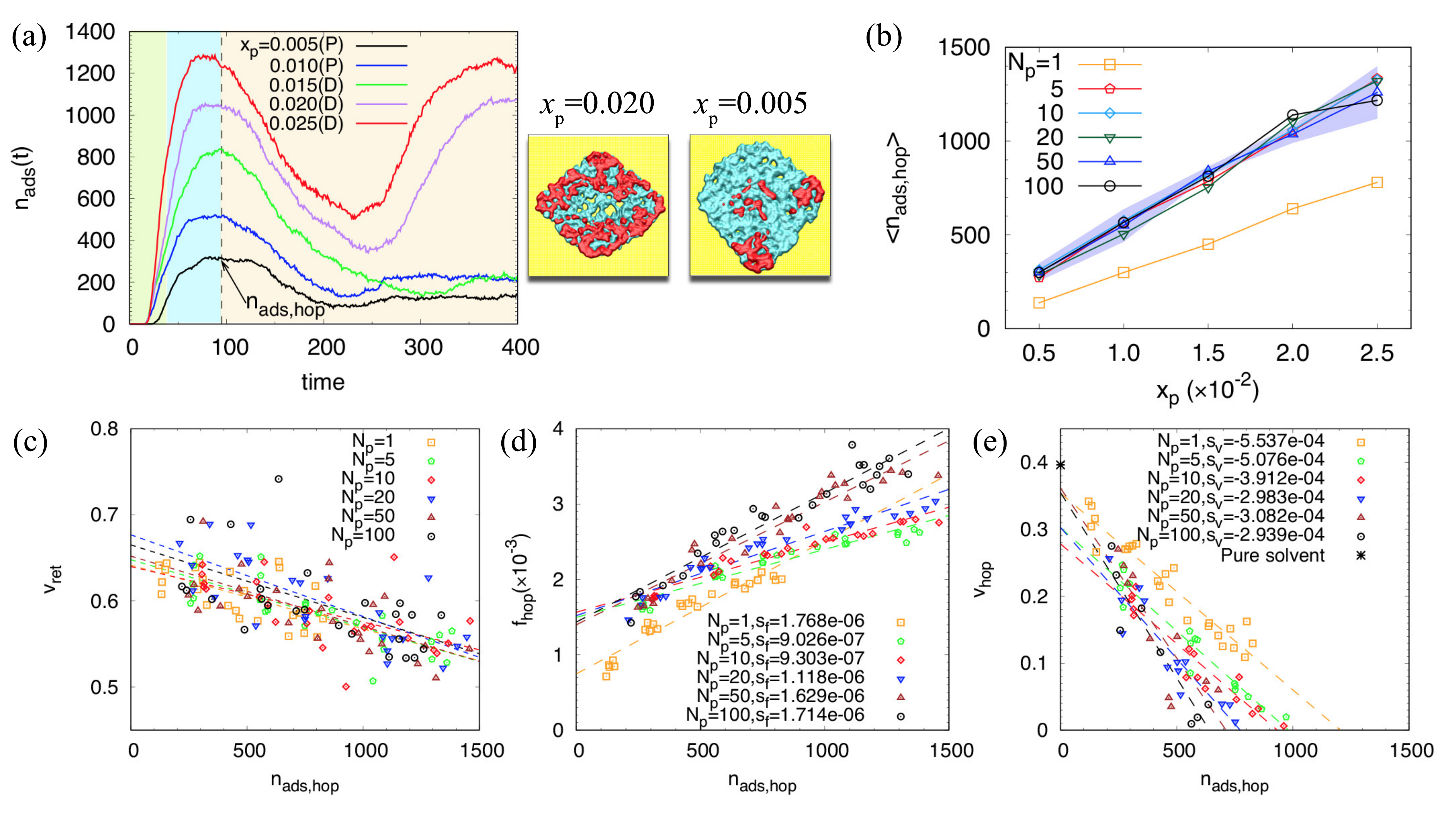}
\caption{\label{fig-padspcomp}(a) The number of adsorbed polymer beads on the surface as a function of time at different concentration for $N_\mathrm{p}$=50. The shaded regions indicate the same as in Figures \ref{fig-polcomp}(b) and (c). Snapshots on the right show adsorbed (red) polymer and (cyan) solvent beads on the surface from the top for $x_\mathrm{p}=0.020$ and 0.005. These snapshots are taken in the middle of the retraction stage ($t=80$). (b) An average of $n_\mathrm{ads}$'s at the beginning of the hopping stage as a function of $x_\mathrm{p}$ at different $N_\mathrm{p}$. The blue shade indicates the error for $N_\mathrm{p}=50$. (c) The retraction velocity, (d) the force against the hopping, and (e) the hopping velocity as a function of $n_\mathrm{ads,hop}$ for all trajectories. In (c), (d), and (e), dashed lines indicate the linear function fitted from each $N_\mathrm{p}$ data. }
\end{figure*}

Figure \ref{fig-padspcomp}(a) shows the time evolution of the number of adsorbed polymer beads, $n_\mathrm{ads}$, for $N_\mathrm{p}=50$ at different polymer concentrations. 
At $A_\mathrm{P/W}=-30$, the adsorbed polymer still behaves as a fluid, thus $n_\mathrm{ads}$ does not differ from $\tilde{n}_\mathrm{ads}$.
The adsorption and desorption rates are known not to be universal but to depend strongly on many factors including the shear rate, the polymer concentration, the polymer length, and the polymer-surface attraction strength.\cite{dutta2013, dutta2015}
Under our specific condition, $n_\mathrm{ads}$ increases during both the spreading and retraction stages, and it starts to decrease at the beginning of the hopping stage.
We define $n_\mathrm{ads}$ at the beginning of the hopping (where $\langle v_\mathrm{axial}\rangle$ in Figure \ref{fig-polcomp}(c) is maximum) as $n_\mathrm{ads,hop}$ to characterize the polymer contribution to the anti-rebound.
Its average from five independent trajectories at each polymer concentration is plotted as a function of $x_\mathrm{p}$ in Figure \ref{fig-padspcomp}(b).
This figure shows that for every polymer length $N_\mathrm{p}$, $n_\mathrm{ads,hop}$ increases linearly, but it hardly depends on $N_\mathrm{p}$ except for the monomeric additive of $N_\mathrm{p}$=1. 

The effect of $n_\mathrm{ads,hop}$ on the rebound tendency can be separated into those on retraction and on hopping.
Simulation snapshots of the adsorbed liquid particles from the top in the middle of the retraction stage (Figure \ref{fig-padspcomp}(a)) shows the adsorbed polymer (red) accumulating close to the contact line, as it retracts together with the contact line.
The concentrated polymer at the contact line reduces the retraction velocity, and the friction gets stronger as more polymer accumulates at the end of retraction. 
To quantify the polymer contribution to the retraction velocity, the retraction velocity of individual trajectories is plotted against $n_\mathrm{ads,hop}$ in Figure \ref{fig-padspcomp}(c).
Here, as expected, we find that a larger polymer adsorption leads to slower retraction.
The slopes of the fitted linear functions for different $N_\mathrm{p}$ do not significantly differ from each other, which means that the contribution of the elongation force for the chosen parameters is small.

At the same time, the adsorbed polymer also strongly affects the hopping motion.
We plot, in Figure \ref{fig-padspcomp}(d), the force acting on the droplet during the hopping stage ($f_\mathrm{hop}$) which is obtained as the slope of a linear fit to the average axial velocity of droplets during the hopping stage in Figure \ref{fig-polcomp}(c).
Figure \ref{fig-padspcomp}(d) shows $f_\mathrm{hop}$ to be linearly correlated with $n_\mathrm{ads,hop}$.
Thus, a larger amount of polymer adsorption leads the surface becoming more wettable by the solvent. 
The elongation force acting on the hopping droplet increases also with increasing $N_\mathrm{p}$, which is seen by the slope of the line increasing with $N_\mathrm{p}$. 
Finally, the hopping velocity ($v_\mathrm{hop}$), the constant velocity of a droplet right after it detaches from the surface (only for rebounding droplets, Figure \ref{fig-padspcomp}(c)), for each trajectory is plotted as a function of $n_\mathrm{ads,hop}$. 
As it adds the polymer contributions to retraction and hopping,  $v_\mathrm{hop}$ is also linearly anti-correlated with $n_\mathrm{ads,hop}$.
The dependence of $v_\mathrm{hop}$ on polymer length can still be identified, the origin of which is the larger elongation force for longer polymer chains.

\subsection{Droplet Size}

\begin{figure}[t!]
\centering
\includegraphics[width=0.48\textwidth]{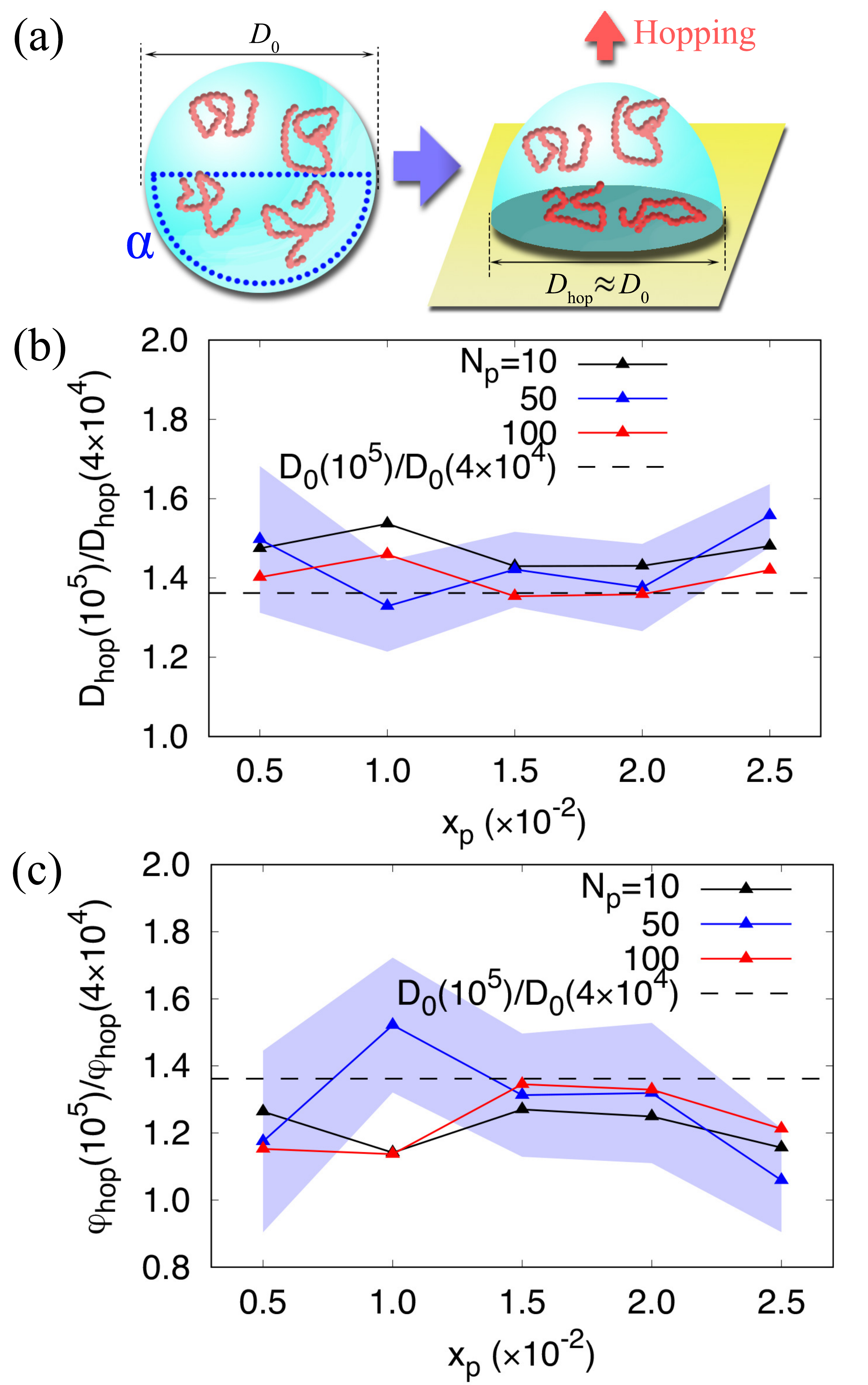}
\caption{\label{fig-sizedep}(a) A schematic of the adsorption process. Pink and red polymer chains indicate unadsorbed and adsorbed polymer, respectively. In this figure, the fraction of a droplet deformation, $\alpha$, is described in the blue dashed line. (b) The ratio of diameter and (c) the ratio of the surface coverage between the large and the small droplets. In (b) and (c), the blue shade shows an error from five trajectories, and the horizontal line shows the ratio of the initial diameter between the large and the small droplets.  }
\end{figure}

One characteristic of the anti-rebound mechanism which is found to be different in the current study ($N=10^5$, $D_0=31.5$) compared to our former one ($N=4\times10^4$, $D_0=23.2$) is the slower retraction velocity.
As a result, the boundary between rebound and deposition slightly shifts toward a lower concentration at any given polymer length (the blue dashed line in Figure \ref{fig-polcomp}(a) indicates the boundary obtained from the smaller droplet.).

To explain the difference, we make a few assumptions, which will be confirmed below.
The first assumption is that polymer has a large enough tendency to adsorb on the surface, so it adsorbs as soon as it touches the surface.
Interchain interaction is also negligible at a small enough polymer concentration.
We consider that a fraction $\alpha$ of the droplet is deformed when a droplet with an initial diameter $D_0$ and the polymer concentration $x_\mathrm{p}$ impacts on a surface with a certain velocity (Figure \ref{fig-sizedep}(a)). 
Polymer molecules which were located in the deformed part of the droplet have a chance to get adsorbed, while others do not.
Then, the number of adsorbed monomer at the beginning of hopping is $n_\mathrm{ads,hop}=\alpha x_\mathrm{p}\pi D_0^3/6$.
We have seen that the rebound tendency is linearly correlated to $n_\mathrm{ads,hop}$.
When droplets of different sizes are compared, however, it is rather related to the surface coverage by adsorbed polymer at the liquid-solid interface, $\phi$.
The surface coverage at the beginning of hopping is given by $\phi_\mathrm{hop}=\frac{n_\mathrm{ads,hop}}{A_\mathrm{hop}}=\frac{\alpha x_\mathrm{p}\pi D_0^3/6}{\pi D_\mathrm{hop}^2/4}$, where $A_\mathrm{hop}$ and $D_\mathrm{hop}$ refer to the area and the diameter of a circle formed by the three-phase contact line at the beginning of hopping, respectively.
Because the hopping stage begins when the droplet is not compressible anymore along parallel to the surface during retraction, we can approximate $D_\mathrm{hop}\approx D_0$. 
Analyzing our trajectories, we find that the ratio of $D_\mathrm{hop}$'s between the large and the small droplets is similar to that of $D_0$'s regardless of $N_\mathrm{p}$ (Figure \ref{fig-sizedep}(b)).
From this, we obtain the relation, $\phi_\mathrm{hop}\sim\alpha x_\mathrm{p} D_0$ for $x_\mathrm{p}<<1$.
Figure \ref{fig-sizedep}(c) shows that the ratio of calculated $\phi_\mathrm{hop}$'s between two droplets of the two sizes is close to that of their $D_0$'s (1.36). 
Increasing the drop size at the same polymer concentration, thus, leads to $\approx 36\%$ denser surface coverage by polymer.
The higher polymer surface density leads to slower contact-line retraction, as has been discussed above. (See Figure \ref{fig-padspcomp}(a)). 
Similarly, more adsorbed polymer chains provide more resistance against hopping.
Thus, the overall tendency to rebound is reduced.

\subsection{Impact Velocity}

\begin{figure}[t!]
\centering
\includegraphics[width=0.48\textwidth]{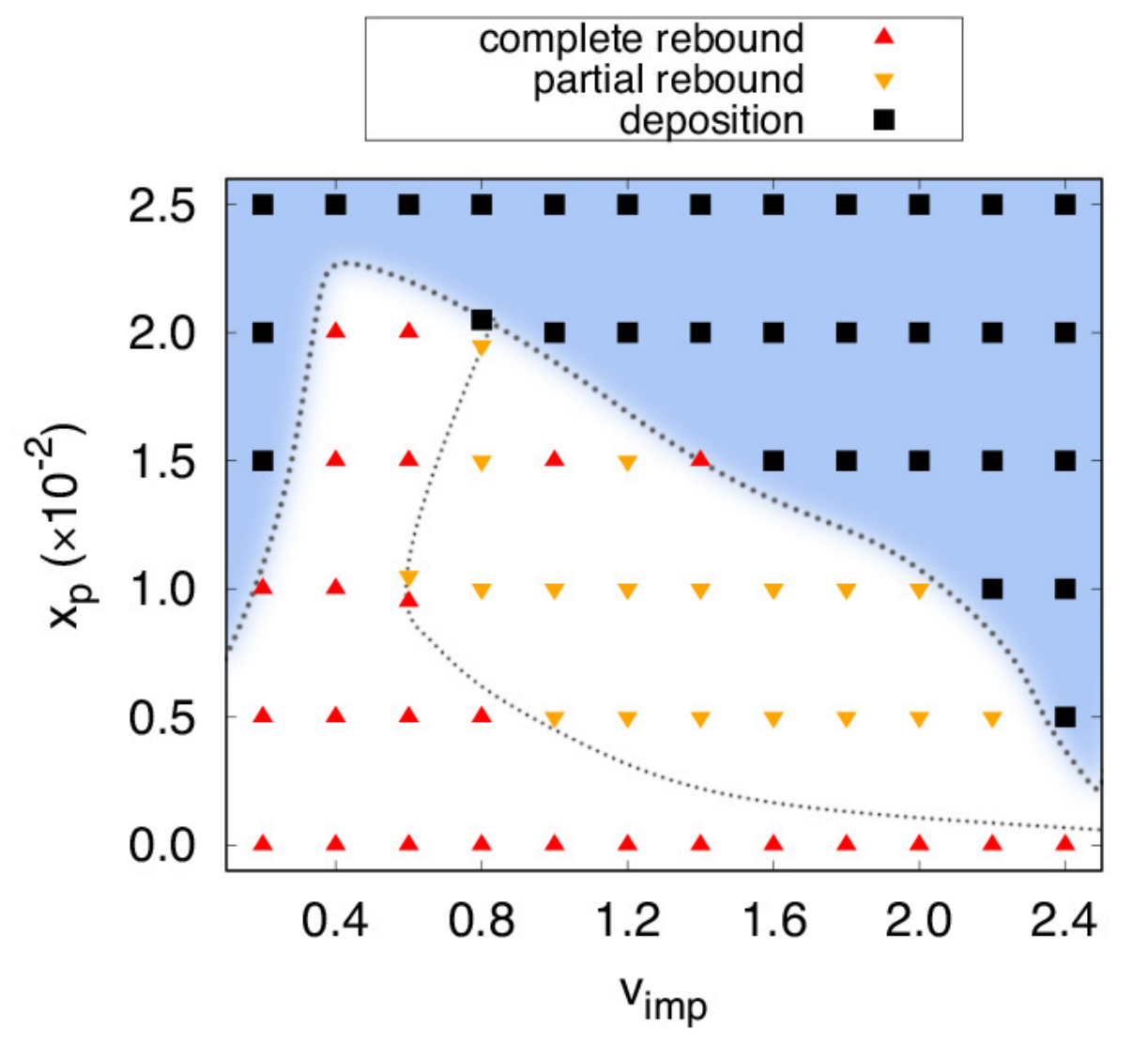}
\caption{\label{fig-diavimp} The most probable outcome of rebound from five independent trajectories for $N_\mathrm{p}=50$ at each polymer concentration and impact velocity. The thick dotted line shows the rough boundary between rebound and deposition, and the thin dotted line indicates the one between complete rebound and partial rebound. The points displaying two outcomes mean that those two are equally probable. }
\end{figure}

We discussed the key role of the adsorbed polymer in the rebound suppression.
This section deals with the dependence of $n_\mathrm{ads,hop}$ on the impact velocity and its effect on the rebound.
As mentioned above, the impact velocity dramatically changes the rebound outcome of Newtonian and non-Newtonian droplets. 
In order to investigate the effect of the impact velocity ($v_\mathrm{imp}$) for droplets with polymer additives, we performed impact simulations for $N_\mathrm{p}$=50 at different polymer concentrations and impact velocities.
The impact velocity ranges from 0.2 to 2.4 in DPD units, which corresponds to We from 1.08 to 156, covering a wide range of a droplet deformation.
Figure \ref{fig-diavimp} shows a diagram of the most probable outcome of rebound obtained from five independent trajectories. 
Complete rebound is mostly observed at low impact velocity. 
Interestingly, we found that the polymer concentration at the rebound-deposition boundary ($x_\mathrm{p}^*$) non-monotonously varies, as $v_\mathrm{imp}$ increases.
While the regime of $x_\mathrm{p}^*$ decreasing with increasing $v_\mathrm{imp}$ (here at large $v_\mathrm{imp}$>0.4) was experimentally observed,\cite{dhar2019} the increase at smaller $v_\mathrm{imp}$ ($< 0.4$)  was not.
Above a certain concentration of $x_\mathrm{p}\geq2.5$, a droplet does not rebound at all. 
Dhar \textit{et al.} explained their experimental observation of $x_\mathrm{p}^*$ decreasing with increasing $v_\mathrm{imp}$ by using the shear rate achieved by the impact velocity which should be faster than the polymer relaxation for the elongation force.\cite{dhar2019}
However, the adsorbed polymer which plays the critical role in the anti-rebound has a much slower relaxation than polymer in solution, so the Weissenberg number alone cannot explain this phenomenon.

\begin{figure}[t!]
\centering
\includegraphics[width=0.9\textwidth]{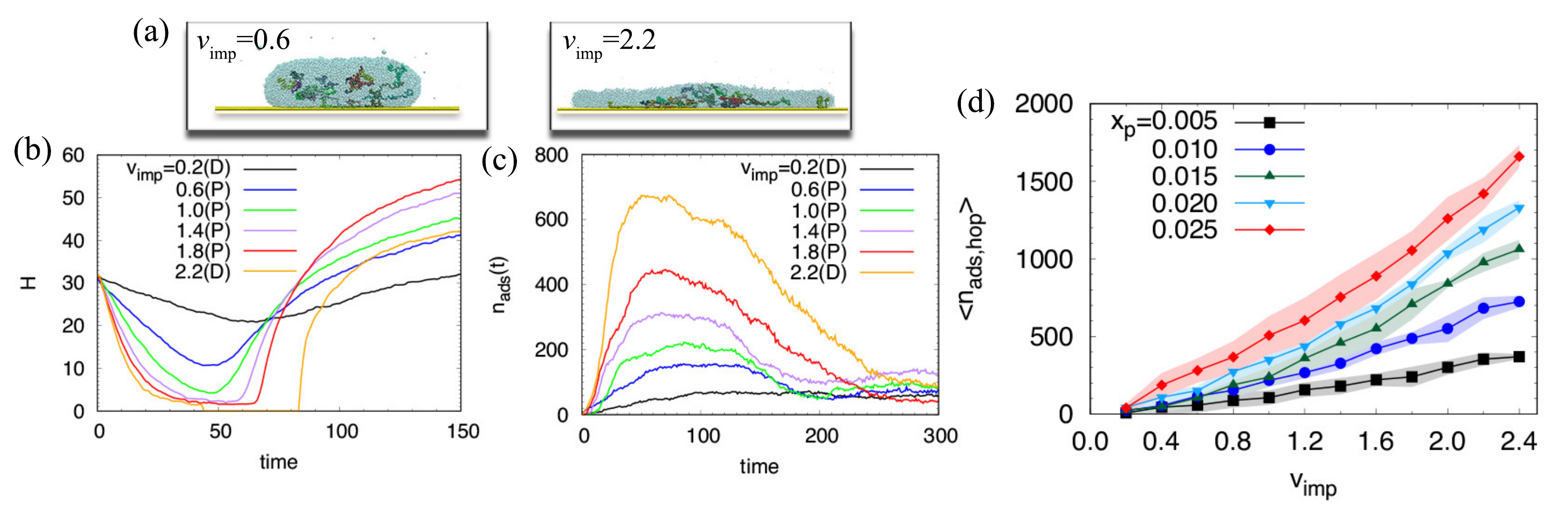}
\caption{\label{fig-nadsvimp}(a) Side views of the simulation snapshots at the maximum spreading for different impact velocities. The color code is the same as used in Figure \ref{fig-outcomes}. (b) The height of the droplet and (c) the number of adsorbed polymer beads as a function of time for different impact velocities.  In (b) and (c), the letter in the legend indicates the outcome as in Figure \ref{fig-polcomp}. (d) The average $n_\mathrm{ads}$ at the beginning of the hopping stage as a function of the impact velocity for different polymer concentrations. The error is shown by the shade.}
\end{figure}

A more straightforward explanation of the rebound outcome in Figure \ref{fig-diavimp} is provided, again, by polymer adsorption.
We find that the number of adsorbed polymer beads significantly depends on $v_\mathrm{imp}$ which drives the droplet deformation during spreading and retraction. 
Comparing the droplet shapes and polymer conformations at maximum spreading for different $v_\mathrm{imp}$ in Figure \ref{fig-nadsvimp}(a), different likelihoods for polymer adsorption are obvious. 
For the small impact velocity of $v_\mathrm{imp}=0.6$, the droplet is not much deformed and the polymer molecules located in the upper part of the polymer do not have the chance to reach the surface. 
In contrast, for $v_\mathrm{imp}=2.2$, most polymer molecules are adsorbed on the surface at the maximum spreading, since the droplet is much flatter and and torus-like.
This is even clearer in the correlation between the height of a droplet ($H$) and the number of adsorbed polymer beads in a droplet as a function of time shown in Figures \ref{fig-nadsvimp}(b) and \ref{fig-nadsvimp}(c).
Here, $H$ is the height of the droplet at its center.
In these figures, anti-correlation between $H$ and $n_\mathrm{ads}$ as a function of time can be easily identified especially during the spreading and the retraction stages. 
Higher $v_\mathrm{imp}$ leads to smaller $H$, which, in turn, results in more polymer adsorption.
This finding also confirms the assumptions that we made about the droplet size effect (Figure \ref{fig-sizedep}(a)).
The average of $n_\mathrm{ads,hop}$ as a function of $v_\mathrm{imp}$ at different polymer concentrations (Figure \ref{fig-nadsvimp}(d)) shows an increase with both $v_\mathrm{imp}$ and $x_\mathrm{p}$ increase.
The almost linear relation of $\langle n_\mathrm{ads,hop}\rangle$ with $x_\mathrm{p}$ indicates that chains adsorb essentially independent of one another at this concentration.

\begin{figure}[t!]
\centering
\includegraphics[width=0.9\textwidth]{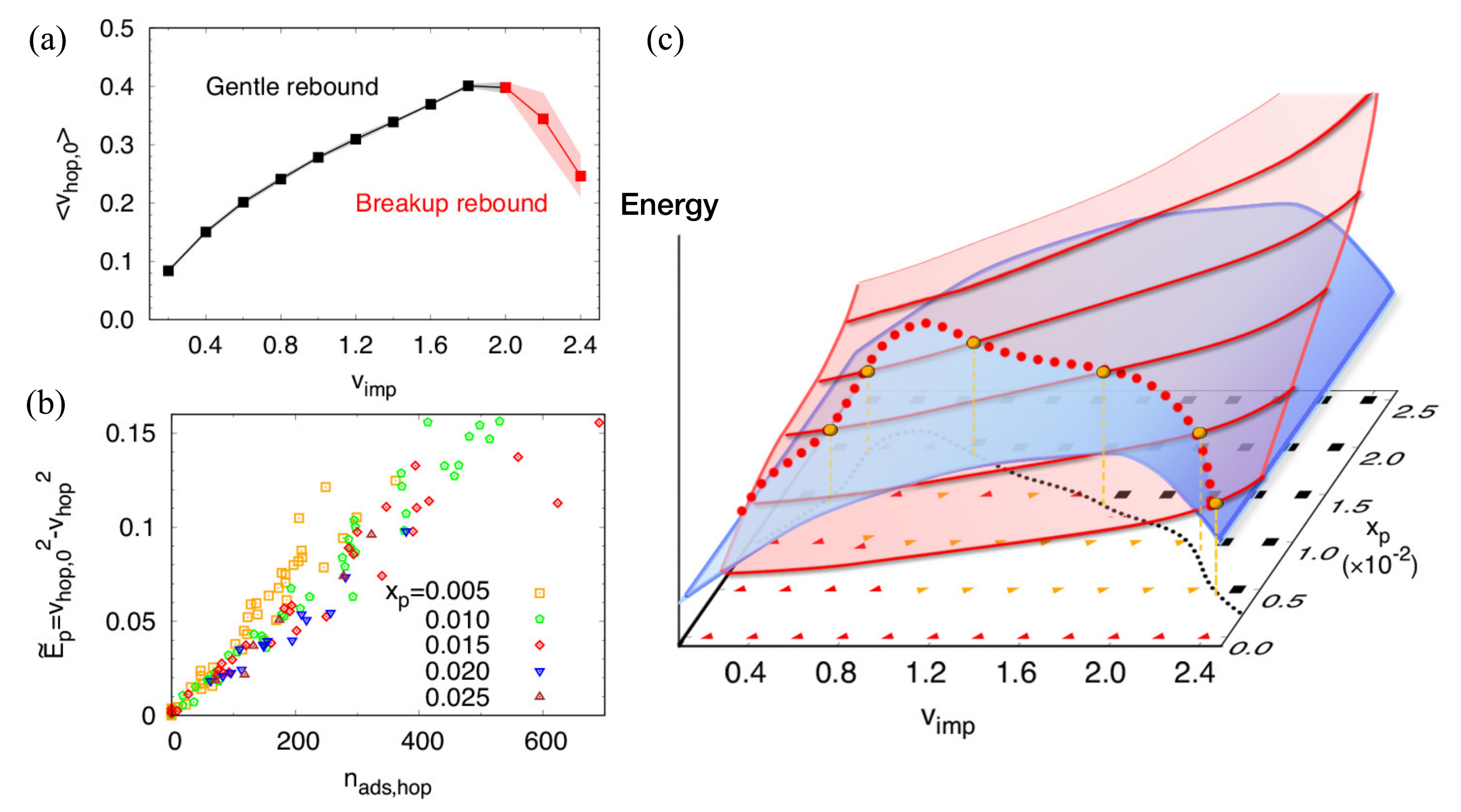}
\caption{\label{fig-polvimp}(a) An average hopping velocity of the pure solvent droplet as a function of the impact velocity. The black and the red points indicate the gentle and the breakup rebounding regimes, respectively. (b) Polymeric energy contribution ($v_\mathrm{hop,0}^2-v_\mathrm{hop}^2$) against the hopping as a function of $n_\mathrm{ads,hop}$ for all rebounding trajectories. (c) A schematic description for (the blue surface) the rebound energy of pure solvent droplets and (red lines and the red surface) the polymeric energy contribution against the rebound.  The resulting boundary between rebound and deposition is determined by the intersection between two surfaces (shown in a red lines on the surfaces and in a black line on the bottom diagram of rebound outcomes. The diagram of rebound outcomes on the bottom is same with one in Figure \ref{fig-diavimp}.} 
\end{figure}

In order to better understand the role of the adsorbed polymer, we need to quantify the polymer contribution to the resistance against the rebound.
For reference, we first calculate $v_\mathrm{hop}$ for the pure solvent droplet at different $v_\mathrm{imp}$ in Figure \ref{fig-polvimp}(a).
In this figure, the hopping velocity of the pure droplet ($v_\mathrm{hop,0}$) is non-monotonous.
It increases with increasing $v_\mathrm{imp}$ for $v_\mathrm{imp}<2.0$, which can be easily understood by the initial kinetic energy. 
However, the opposite behavior is found for $v_\mathrm{imp}>2.0$.
In this regime of high $v_\mathrm{imp}$, a droplet rebounds  with fragmentation or distortion, which is typically called a breakup rebound.\cite{yarin2006}(See SI movies)
Fragmentation happens during retraction.
Therefore, the energy is additionally dissipated by the splashing tiny droplets, and is not available for the hopping process of the mother droplet.
If the droplet is highly distorted during spreading and retraction, the contact line retracts asymmetrically and a part of the energy is stored in the shape oscillations after retraction, which also decreases the hopping velocity.
Using the hopping velocity $v_\mathrm{hop,0}$ of the pure droplet as a reference, we calculate $\tilde{E}_\mathrm{p}\equiv v_\mathrm{hop,0}^2-v_\mathrm{hop}^2$ which corresponds to the polymer contribution to the energy dissipation.
Figure \ref{fig-polvimp}(b) shows $\tilde{E}_\mathrm{p}$ as a function of $n_\mathrm{ads,hop}$ for all trajectories of gently rebounding droplets.
In this figure, $\tilde{E}_\mathrm{p}$ is linearly proportional to $n_\mathrm{ads,hop}$ for each $x_\mathrm{p}$.
The slope of the linear fit for each $x_\mathrm{p}$ seems to only weakly depend on $x_\mathrm{p}$.
This means that we can correlate the polymeric contribution to the additional energy dissipation during the rebound linearly with $n_\mathrm{ads,hop}$. 

Finally, the diagram of the rebound outcome in Figure \ref{fig-diavimp} can be understood with Figure \ref{fig-polvimp}(c).
A droplet of pure solvent has a rebound energy proportional to $\langle v_\mathrm{hop,0}^2 \rangle$, that is shown by the blue surface in this figure.
If polymer is added to a droplet with a certain concentration $x_\mathrm{p}$, energy is additionally dissipated, the amount of which, $\tilde{E}_\mathrm{p}$, is proportional to the number of adsorbed polymer beads.
The average adsorbed amount, $\langle n_\mathrm{ads,hop}\rangle$, as a function of concentration and of the impact velocity in Figure \ref{fig-nadsvimp}(c), therefore, directly determines the amount of dissipated energy by polymer.
This polymer contribution is represented by the red surface (the red lines correspond to constant concentrations.) in Figure \ref{fig-polvimp}(c).
If the pure-solvent rebound energy (blue surface) is larger than the energy dissipation due to polymer additive (red surface), the excess rebound energy is converted into the hopping velocity.
Otherwise, the polymer successfully suppresses the droplet rebound by means of the additional friction during retraction and the resistance against hopping. 
The intersection between the blue and the red surfaces comprises the rebound-deposition boundary on the $x_\mathrm{p}$-$v_\mathrm{imp}$ space (the red dashed line indicating the intersection of the surfaces projected on the bottom as the black dashed line). 
This interpretation also explains well the unexpected boundary at low $v_\mathrm{imp}$ by the fact that the rebound energy even without polymer is very low, whence already a small amount of adsorbed polymer suppresses the rebound.

\section{Conclusion}

In this work, we investigated the mechanism of the droplet rebound suppression by a small amount of polymer additive when a droplet impacts on a solvophobic surface. 
The key feature is the adsorption or partial adsorption of the polymer into the surface.
It slows the retraction and it impedes the hopping motion.
The adsorbed polymer mediates an additional attraction between the solvent and the surface, which creates additional friction during retraction and the adsorptive force during hopping.
The polymers are highly stretched following the flow of liquid, thus, pull a droplet during hopping.
The relative importance of the two mechanisms depends on the polymer-surface attraction. 
If it is weak, there mainly is resistance against hopping.  
For strong attraction, the reduction of the retraction velocity is the main impediment to rebound.
When it is too strong, polymer is irreversibly adsorbed on the surface, and rebound is no longer suppressed due to the lack of elongation force.
In agreement with our previous study, both a larger polymer concentration and a longer polymer chains increase the rebound suppression.
Increasing droplet size results in more situations where both mechanisms are simultaneously active.
We expect that for macroscopic droplets, rebound suppression is achieved by both mechanisms. 
Faster impacting droplets can, in general, be prevented from rebounding by small polymer content than slower droplets due to their larger energy dissipation by the adsorbed polymer. 
However, also for very small impact velocities of We$\approx$1 when the droplet is hardly deformed, rebound is less likely.
This is attributed to the fact that the energy of pure solvent without polymer is already very small, thus even a very small amount of adsorbed polymer is able to suppress the rebound.
This work provides a clear molecular picture of the anti-rebound mechanism, which can address the unanswered questions from existing experiments on rebound suppression by polymer additives. 
When combined with further investigation of polymer adsorptivity under different conditions, e.g., adsorption strength and shear rate, this work will also open up the possibility of a more sophisticated control of droplet rebound.

\begin{acknowledgements}
We would like to thank for the financial support by the Collaborative Research Centre Transregio 75 (project number 84292822) 
\textit{Droplet Dynamics in Extreme Environments} of the Deutsche Forschungsgemeinschaft. Calculations for this research were conducted on the Lichtenberg high performance computer of the TU Darmstadt and the GCS Supercomputer JUWELS at J\"ulich Supercomputing Centre (JSC) through the John von Neumann Institute for Computing (NIC).
\end{acknowledgements}


\bibliography{droplet}

\end{document}